# Optical Neutrality: Invisibility without Cloaking

Reed Hodges,[1] Cleon Dean,[1] Maxim Durach[1],*

[1]*Department of Physics, Georgia Southern University, 65 Georgia Avenue Math/Physics Bldg., Statesboro, GA 30460*
*\*Corresponding author: mdurach@georgiasouthern.edu*

**We show that it is possible to design an invisible wavelength-sized metal-dielectric metamaterial object without evoking cloaking. Our approach is an extension of the neutral inclusion concept by Zhou and Hu [Phys.Rev.E 74, 026607 (2006)] to Mie scatterers. We demonstrate that an increase of metal fraction in the metamaterial leads to a transition from dielectric-like to metal-like scattering, which proceeds through invisibility or optical neutrality of the scatterer. Formally this is due to cancellation of multiple scattering orders, similarly to plasmonic cloaking introduced by Alu and Engheta [Phys.Rev.E 72, 016623 (2005)], but without introduction of the separation of the scatterer into cloak and hidden regions.**

*OCIS codes:* (290.5839) Scattering, invisibility; (290.4020) Mie theory; (160.3918) Metamaterials; (250.5403) Plasmonics.

Most objects scatter light under normal circumstances. Subwavelength objects are characterized by Rayleigh scattering, whereas larger particles exhibit Mie scattering with predominant forward scattering [1-3]. From a materials perspective dielectrics tend to transmit light in the forward direction, whereas metals act mostly as opaque objects, which reflect or back-scatter light and form shadows [1-3]. Invisible objects, which do not scatter, have been a Holy Grail of photonics [4]. Over the course of the recent decade invisible photonic structures has grown into a major branch of photonics research. In transformation optics cloaking, the cloak serves to guide rays around the hidden object, which is made invisible to rays and is supposed to be large with respect to the radiation wavelength [5-7]. In plasmonic cloaking, cloaks are shells wrapped around hidden objects, which are designed to cancel the dominant multipole orders of radiation, scattered from sub-wavelength and wavelength-sized objects [8-11]. In all cases invisibility is achieved via introduction of a cloak which surrounds the volume to be hidden. One departure from this principle is the proposal of using complementary media to induce invisibility by a cloak, which does not enclose the object to be hidden [12]. Another interesting development of the plasmonic cloaking idea is mantle cloaking, in which surface currents running at a specially designed metasurface, enclosing the hidden object, cancel scattering from the object [13-14]

An alternative interpretation of invisibility is based on the "neutral inclusion" idea, where the invisibility conditions are understood as the quasistatic effective permittivity being unity [15, 16]. Recently, it has been proposed that a sub-wavelength homogeneous sphere can be made invisible, without evoking cloaking [17-19]. This was demonstrated for a radially symmetric anisotropic metamaterial sphere in the Rayleigh approximation, where the quasistatic effective permittivity of the sphere becomes equal to unity and the induced dipole is suppressed. There is a strong interest in extending such optical neutrality and transparency to mesoscale metal-dielectric Mie scatterers [20]. In this paper using full Mie theory we show that formally this quasistatic mechanism of invisibility without cloaking is intimately related to the plasmonic cloaking and can be extended to design wavelength-sized invisible homogenized objects, such that a solitary structure, which cannot be separated into a hidden object/cloak pair, can still be designed to be invisible, i.e. exhibiting *optical neutrality*. Moreover, we show that such optical neutrality can be thought of as a transitional phase of transformation of a metal-dielectric metamaterial from a dielectric into a metal upon increase of the metal fraction.

We consider spheres with radial anisotropy, whose dielectric tensor $\hat{\varepsilon} = \varepsilon_r \hat{r}\hat{r} + \varepsilon_t(\hat{\theta}\hat{\theta} + \hat{\phi}\hat{\phi})$ has different diagonal elements $\varepsilon_r$ and $\varepsilon_t$ in radial and in angular directions. We use the effective medium permittivities for layered media [21] $\varepsilon_r^{-1} = \varepsilon_m^{-1} f + \varepsilon_d^{-1}(1-f)$ and $\varepsilon_t = \varepsilon_m f + \varepsilon_d(1-f)$, where $\varepsilon_m$ is the dielectric permittivity of metal, taken to be gold in our study [22], and $\varepsilon_d = n_d^2$ is the permittivity of the dielectric component. We denote the volumetric fraction of metal as $f$. The spheres we consider do not exhibit magnetic response. Such structures serve as an effective medium approximation to layered spheres, composed of subwavelength concentric metal and dielectric layers in an onion-like fashion, also known as matryoshka nanoshells [23, 24]. Radially anisotropic particles have attracted considerable attention due to their scattering [25-27], light-trapping [28] and strong-coupling [29] properties. The electromagnetic fields scattered by the metamaterial sphere as well as the scattering cross-sections can be described with a framework of Mie theory [25]. The main difference from the scattering on isotropic spheres is the unconventional order number $\nu = [(l(l+1)\varepsilon_t/\varepsilon_r + 1/4)^{-1/2} - 1/2]$ which characterizes the $l$th vector spherical harmonics of TM polarized fields and depends on the electric anisotropy ratio $\varepsilon_t/\varepsilon_r$.

In Fig. 1 we show scattering from a dielectric sphere with refractive index $n_d = 1.33$, a gold sphere, and a metal-dielectric metamaterial sphere, all with the same radius $a = 350$ nm at $\lambda = 700$ nm. One can see that the dielectric sphere (Fig. 1(a)) scatters predominately in the forward direction and forms a photonic jet [30-31]. The metal sphere exhibits both forward and backward scattering due to its strong plasmonic dipole response (Fig. 1(b)) [32-34]. For comparison, we show scattering of light on the metamaterial sphere with metal fraction $f = 5\%$ (Fig. 1(c)), which shows only mild near-field scattering and complete recovery of the incident wavefronts in the shadow the sphere. This serves as an example of the optical neutrality proposed in this paper.

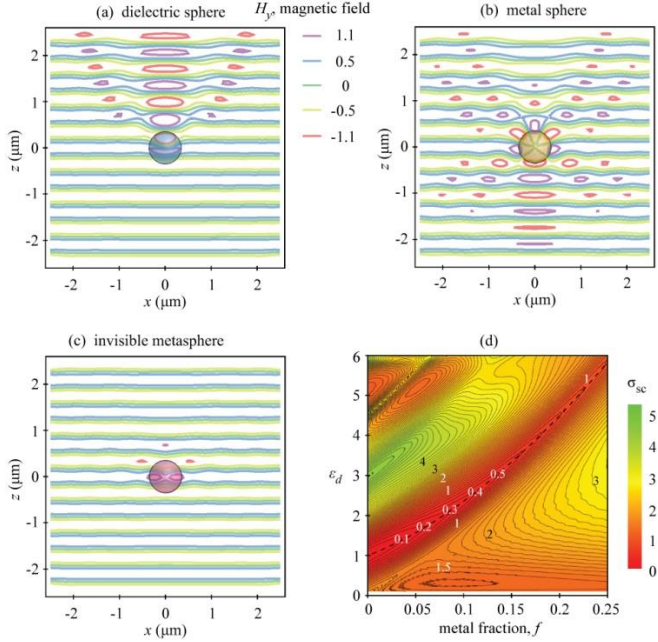

Fig. 1. Magnetic field distribution snapshots for 350-nm radius spheres. These contour plots exhibit the transition from a photonic nanojet-producing dielectric sphere (a), to a plasmonic metal sphere (b), to an invisible metal-dielectric metamaterial sphere (c). Panel (d) shows a contour plot of the normalized scattering cross section as a function of dielectric permittivity of the dielectric inclusion and the metal (gold) fraction $f$. The dashed line shows the parameters for the optical neutrality.

To illustrate the idea of the "neutral inclusion" for wavelength sized objects in Fig. 1(d) we plot the dependence of the normalized scattering cross-section $\sigma_{sc}$ of a 700 nm diameter sphere at $\lambda = 700$ nm as a function of $\varepsilon_d$ and metal fraction $f$. In this graph red areas indicate the reduced $\sigma_{sc}$ as indicated by the labels on the contours. For an inherently "neutral" empty spherical region with $\varepsilon_d = 1$ and $f = 0$ there is no scattering, but if one fills this volume with just 5% of metal, keeping the rest empty $\sigma_{sc}$ becomes ~1. Alternatively, if one fully fills the spherical region with a low-index $n_d = 1.33$ dielectric, the cross section becomes $\sigma_{sc} \approx 1.93$. Nevertheless, if one introduces the dielectric and metal into the volume of the sphere following the dashed black line in Fig. 1(d) the cross section increases much slower than along other directions of neutrality breaking, in essence preserving neutrality.

To get a better grasp of the predicted neutrality effect consider an object exposed to action of an external electromagnetic field, with electric field vector $\boldsymbol{E}_{ext}(\boldsymbol{r}, t)$. In response, the object generates a scattered field $\boldsymbol{E}_{sc}(\boldsymbol{r}, t)$. Internally a polarization is generated $\boldsymbol{P}_{int}(\boldsymbol{r}, t)$ and the object permits penetration of the internal field $\boldsymbol{E}_{int}(\boldsymbol{r}, t)$. The continuity of the tangential components of the electric field at the outermost extremity of the object requires that

$$E_{ext}^{tan}(\boldsymbol{r}_s, t) - E_{int}^{tan}(\boldsymbol{r}_s, t) = E_{sc}^{tan}(\boldsymbol{r}_s, t) \quad (1)$$

where vector $\boldsymbol{r}_s$ is the radius of the outer surface of the object. Simultaneously, due to the polarization of the object a surface charge $\sigma(\boldsymbol{r}_s, t)$ is generated, which is equal to the normal component of the polarization at the outer interface $\sigma = P_{int}^n(\boldsymbol{r}_s, t)$. The continuity of the normal component of the displacement field requires that

$$\sigma(\boldsymbol{r}_s, t) = \frac{E_{ext}^n(\boldsymbol{r}_s, t) - E_{int}^n(\boldsymbol{r}_s, t)}{4\pi} + \frac{E_{sc}^n(\boldsymbol{r}_s, t)}{4\pi} \quad (2)$$

Neutrality implies that scattered fields are negligible $E_{sc}^{tan}(\boldsymbol{r}_s, t) \approx 0$ and $E_{sc}^n(\boldsymbol{r}_s, t) \approx 0$, so that the tangential component of the external excitation field $E_{ext}^{tan}$ is matched by the internal field $E_{int}^{tan}$ at the interface and the surface charge is purely due to the difference between the normal components of internal and external electric fields. Note that this absence of scattering is not based on the introduction of an additional layer or mantle to the object, which cloaks it, but is due to the peculiarity of the *polarization of the object itself*.

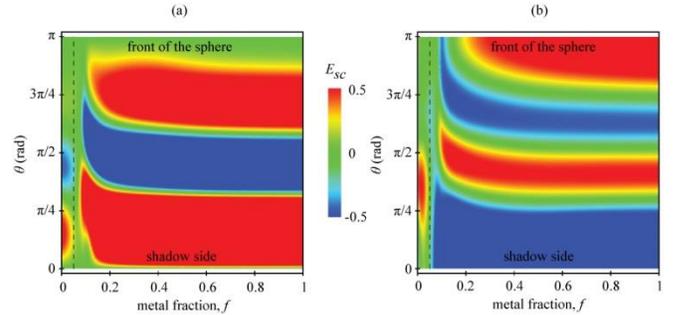

Fig. 2. Snapshots of surface distribution of scattered fields $E_{sc}^n$ (a) and $E_{sc}^{tan}$ (b) for different metal fractions $f$ in an $a = 350$ nm sphere at $\lambda = 700$ nm. The dashed line indicates the transition from the dielectric-like to the metal-like scattering via invisibility at $f = 5\%$.

We plot the scattered fields at the interface of the metasphere in Fig. 2. As one can see for a dielectric sphere $f = 0$ the scattered fields are concentrated on the front side of the sphere and feature only one or two anti-nodes. The scattered fields on the surface of the metal sphere are distributed around the whole sphere and have 3 to 4 anti-nodes. Such metal-like scattering is exhibited by metaspheres in the range of metal fractions from $f \approx 0.1$ to 1. The neutrality condition, which is indicated by the dashed lines at $f = 0.05$, can therefore be seen as a transition between dielectric-like to metal-like scattering optical phases. Metamaterial objects are known for demonstrating various optical phases and responses and transitions between the phases depending on the parameters of the structures [35]. The possibility of considering invisibility as a transition between the dielectric-like and metal-like scattering properties as is proposed in this paper is very promising for future studies in the field of metamaterials.

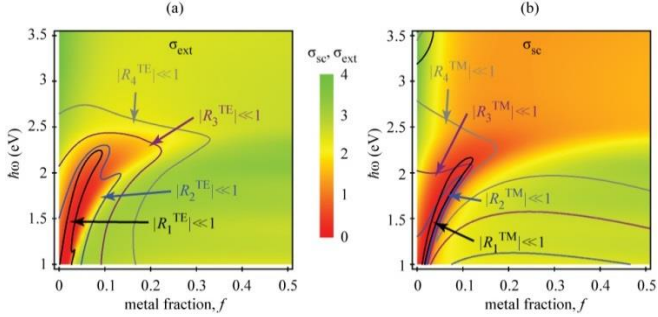

Fig. 3. Density plots of the extinction cross section (a) and the scattering cross section (b) for a range of metal fractions and incident light energies. The contours indicate where the coefficients R approach zero.

To study the neutrality transition in detail we represent the Mie scattering coefficients for the radially anisotropic spheres as

$$b_l = \frac{-R_l^{TE}}{R_l^{TE} + iS_l^{TE}}, \quad a_{l,\nu} = \frac{-R_{l,\nu}^{TM}}{R_{l,\nu}^{TM} + iS_{l,\nu}^{TM}} \quad (3)$$

where the parameters $R$ and $S$, corresponding to multipole order $l$ and $\nu$, are provided in the Supplementary materials. In the quasistatic approximation the scattering cross-section of the radially anisotropic spheres is proportional to the polarizability $\alpha = a^3(\varepsilon_{eff} - 1)/(\varepsilon_{eff} + 2)$ with an effective dielectric permittivity $\varepsilon_{eff} = \varepsilon_r \nu$ [17-19]. Correspondingly, the neutrality condition for radially anisotropic spheres with extremely small radii, $a \ll \lambda$, is $\varepsilon_{eff} = 1$ [16] or equivalently $\varepsilon_r = (2\varepsilon_t - 1)^{-1}$ [17-19]. The neutrality of larger spheres is achieved when the scattering coefficients (Eq. 3) are zero up to multipole order $l \approx 2\pi a/\lambda$. The corresponding coefficients $R$ for these multipoles should vanish. We show the contours of $R_{l,\nu}^{TE,TM} < 0.1$ in Fig. 3 for $l$ from 1 to 4 (higher multipoles, which are not shown for simplicity, vanish as well). As one can see these coefficients are simultaneously small in Mie scattering regime under consideration.

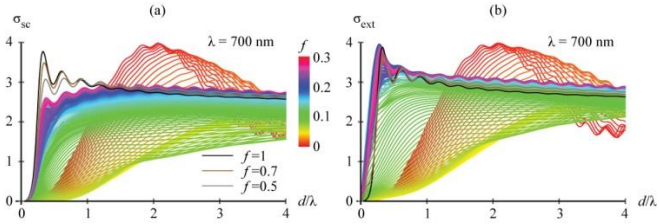

Fig. 4. Plot of the scattering (a) and extinction (b) cross sections as a function of the ratio of sphere diameter to incident wavelength. The cross sections stay very close to zero up to a single-wavelength sized sphere.

In Fig. 3 we show extinction and scattering cross section spectra $\sigma_{ext}$ and $\sigma_{sc}$ for metamaterial spheres with radius $a = 350$ nm for different metal fractions. The red regions in these figures designate areas of negligible cross sections $\sigma \approx 0$. In particular, for $\lambda = 2a = 700$ nm the spheres are invisible for metal fractions $f \approx 0.05$. The regions of neutrality are consistent with the condition that the values of $R$ coefficients being less than 0.1. This means that neutrality transition in a wavelength-sized uncloaked object is due to the simultaneous vanishing of $R$ coefficients for several multipoles.

The maximum size of the sphere which can be made invisible correlates with the number of multipoles for which the R coefficient is zero. We show scattering and extinction cross sections at $\lambda = 700$ nm for different diameters $d = 2a$ and metal fractions $f$ in Fig. 4. For a metal fraction of $f = 0.05$, both the scattering and extinction cross sections are practically zero up to $d \approx \lambda$. They are still reduced three to four times compared to pure dielectric or metal spheres through a diameter twice the incident wavelength; there, they are approximately equal to the geometrical cross section. The metamaterial spheres with $f = 0.05$ to $0.1$ exhibit reduced cross sections for diameters up to several wavelengths (see Fig. 4).

It is interesting to note that the recovery of the cross-sections to the pure metal/dielectric level at $\frac{d}{\lambda} \approx 3$ to $4$ is purely due to increase in the forward scattering in the metamaterial spheres, while they still stay invisible for backscattering measurements as is shown in Fig. 5, where the full scattering cross-section $\sigma_{sc}$ is split into the forward $\sigma_{forw}$ (i.e. $\theta = 0°$ to $45°$) and backward $\sigma_{back}$ ($\theta = 45°$ to $90°$) at $\lambda = 700$ nm. For subwavelength spheres one can observe scattering in the Rayleigh regime with forward and backward scattering similar to each other in magnitude (see solid curve in Fig. 5 (a) for $a = 100$ nm). For larger spheres in the Mie regime the forward scattering dominates, but as was described above both are small at the neutrality transition for $d$ below $\lambda$ (see the dashed curve in Fig. 5 (a) for $a = 200$ nm and the solid curve in Fig. 5 (b) for $a = 350$ nm). When eventually the dimensions of the spheres become larger than wavelength the full cross sections stop being negligible compared to the geometrical sizes, but this happens exclusively due to the increase in the forward scattering, while the backscattering remains negligible as is shown by the solid curves in Fig. 5(b) for $d = 4\lambda = 2.8$ μm.

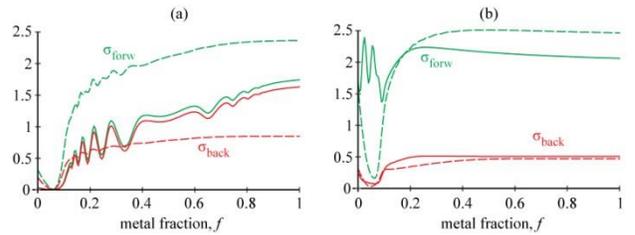

Fig. 5. The forward (green) and backward (red) scattering cross sections for spheres with (a) $a = 100$ nm (solid), $a = 200$ nm (dashed) and (b) $a = 350$ nm (dashed), $a = 1400$ nm (solid).

In conclusion, we have considered the possibility of understanding the invisibility of wavelength-sized object without evoking cloaking, based on the neutral inclusion principle. We have demonstrated that it is possible to design metamaterial metal-dielectric structures with negligible cross-sections, whose invisibility is not due to introduction of a cloak, but stems from the transition of the scattering properties of the structures from dielectric-like to metal-like as the metal fraction is increased.

**Supplementary material.** (a) Scattering coefficients for the radially anisotropic spheres can be expressed using Eq. (1), where

$$R_l^{TE} = j_l(k_0\sqrt{\varepsilon_t}r)\frac{\partial}{\partial r}[rj_l(k_0r)] - j_l(k_0r)\frac{\partial}{\partial r}[rj_l(k_0\sqrt{\varepsilon_t}r)]$$

$$S_l^{TE} = j_l(k_0\sqrt{\varepsilon_t}r)\frac{\partial}{\partial r}[ry_l(k_0r)] - y_l(k_0r)\frac{\partial}{\partial r}[rj_l(k_0\sqrt{\varepsilon_t}r)]$$

$$R_{l,\nu}^{TM} = j_\nu(k_0\sqrt{\varepsilon_t}r)\frac{\partial}{\partial r}[rj_l(k_0r)] - j_l(k_0r)\frac{1}{\varepsilon_t}\frac{\partial}{\partial r}[rj_\nu(k_0\sqrt{\varepsilon_t}r)]$$

$$S_{l,\nu}^{TM} = j_\nu(k_0\sqrt{\varepsilon_t}r)\frac{\partial}{\partial r}[ry_l(k_0r)] - y_l(k_0r)\frac{1}{\varepsilon_t}\frac{\partial}{\partial r}[rj_\nu(k_0\sqrt{\varepsilon_t}r)]$$

**Funding.** Blue Waters - National Science Foundation (ACI 1238993)

**Acknowledgment**. R. H. and M. D. are grateful for the support of the Blue Waters Student Internship Program and the student research grant from College Office of Undergraduate Research (COUR) at Georgia Southern University. M. D. is supported by the Office of the Vice President for Research & Economic Development at Georgia Southern University.